\documentclass[conference]{IEEEtran}
\IEEEoverridecommandlockouts

\usepackage{cite}
\usepackage{amsmath,amssymb,amsfonts}
\usepackage{algorithmic}
\usepackage{graphicx}
\usepackage{textcomp}
\usepackage{xcolor}

\usepackage{multirow}
\usepackage{bbold}
\usepackage{amsfonts} 
\usepackage{amsmath}
\usepackage{hyperref}
\usepackage{svg}

\newcommand{\orcid}[1]{\href{https://orcid.org/#1}{\includegraphics[width=10pt]{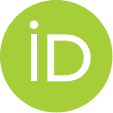}}}

\def\BibTeX{{\rm B\kern-.05em{\sc i\kern-.025em b}\kern-.08em
    T\kern-.1667em\lower.7ex\hbox{E}\kern-.125emX}}
\begin{document}

\title{Efficient Multi-Cycle Folded Integer Multipliers\\
}

\author{
\IEEEauthorblockN{Ahmad Houraniah \orcid{0000-0001-7925-8888}}
\IEEEauthorblockA{\textit{Dept. of  Computer Science} \\
\textit{Ozyegin University}\\
Istanbul, Turkiye \\
ahmad.houraniah@ozu.edu.tr}
\and
\IEEEauthorblockN{H. Fatih Ugurdag \orcid{0000-0002-6256-0850}}
\IEEEauthorblockA{\textit{ Dept. of Electrical and Electronics Engineering} \\
\textit{Ozyegin University}\\
Istanbul, Turkiye \\
fatih.ugurdag@ozyegin.edu.tr}
\and
\IEEEauthorblockN{C. Emre Dedeagac \orcid{0000-0001-7197-9541}}
\IEEEauthorblockA{\textit{Dept. of  Computer Science }\\
\textit{Ozyegin University}\\
Istanbul, Turkiye \\
emre.dedeagac@ozyegin.edu.tr}
}

\maketitle

\begin{abstract}
Fast combinational multipliers with large bit widths can occupy significant silicon area, which also drives up power consumption. Area can be reduced through resource sharing (i.e., folding) at the expense of lower throughput, which is acceptable for some applications. This work explores multiple architectures for Multi-Cycle folded Integer Multiplier (MCIM) designs, which are based on Schoolbook and Karatsuba approaches. Applications sometimes require a fractional number of multiplications to be performed per cycle. For example, an algorithm may only require 3.5 multiplications per cycle. In such a case, 3 multipliers with a throughput of 1 plus an additional smaller multiplier with a throughput of $1/2$ would be sufficient to maintain the algorithm's throughput. Our MCIM design generator offers customization in terms of throughput, latency, and clock frequency. MCIM designs were synthesized and verified for various parameter values using scripts. ASIC synthesis results show that MCIM designs with a throughput of $1/2$ offer area savings of up to 44\% for bit widths of 8 to 128 with respect to directly synthesizing the * operator. Additionally, MCIM designs can offer up to 33\% energy savings and 65\% average peak power reduction.

\end{abstract}


\section{Introduction}
Integer multipliers are essential building blocks in ASICs and CPUs, enabling critical operations in cryptography, signal processing, and machine learning. Emerging applications have shown an increased demand for large integer multiplications \cite{cuda} (e.g., 128b+ for homomorphic encryption). Traditional combinational integer multipliers with large bit width face prohibitive area and power costs, especially for area-constrained applications such as edge or IoT devices, which can produce critical design bottlenecks.

Current solutions for area reduction of multipliers often prioritize FPGA implementations or focus on single-cycle multipliers (or the pipelined counterpart). The prior work lacks ASIC-optimized multi-cycle multipliers that balance area efficiency and performance. FPGA-based implementations target efficient utilization of FPGA primitives (e.g., DSPs), which poorly exploit ASIC customization.

In this paper, we propose Multi-Cycle Folded Integer Multipliers (MCIMs), which are ASIC-focused RTL generators with compile-time tunable throughput, latency, and area tradeoffs. 

Through synthesis-driven evaluations (on TSMC 40nm) for a range of timing constraints and multiplication sizes, we demonstrate the area and power savings that can be achieved by MCIMs. The major contributions of this work are:

\begin{enumerate}
    \item First ASIC-focused exploration of multi-cycle folded multipliers, offering up to 33\% energy savings and 65\% average peak power reduction, as well as area savings up to 44\% (for cycle-time=2).

    \item Fractional throughput architectures and design generators, enabling highly customized multipliers (e.g., 5/6 throughput).
    
    \item Recursive Karatsuba-based multipliers offering 64\% area savings for multiplications with bit widths of $ 128 \times 128$.

\end{enumerate}

The rest of the paper is organized as follows. Section \ref{section:previous work} presents the previous work. Section \ref{proposed architectures} presents the architectures proposed in this work. Section \ref{results} presents the implementation results, while Section \ref{conclusion} concludes the paper.

\section{Previous Work}\label{section:previous work}

Whether a multiplier is multi-cycle or single-cycle, its underlying algorithm should be efficient to start with. In \cite{wallace1964suggestion}, Wallace's carry-save addition revolutionized partial product reduction by compressing sums of three numbers into two, forming the basis for modern carry-save trees. 

This is especially important for large integer multiplications because of quadratic area growth with bit width, necessitating divide-and-conquer strategies to balance efficiency and scalability. The Schoolbook approach, derived from the distributive property of multiplication, can be utilized for area reduction, where a smaller multiplier is used repeatedly for the computation (i.e., resource sharing). Using Schoolbook multiplication, the complexity for multiplication is ${\Theta }(n^2)$. Karatsuba \cite{karatsuba1962multiplication} proposed an algorithm that reduces the complexity to ${\Theta }(n^{1.59})$. The Karatsuba algorithm can be described by equations \ref{eq:karatsuba_1level_1}, \ref{eq:karatsuba_1level_2}, and \ref{eq:karatsuba_1level_3} for a single level of division, where the number of smaller multiplications required is reduced from 4 to 3.
\\
\begin{equation}\label{eq:karatsuba_1level_1}
T_1 = A_1 \times B_1, \hspace{4pt} T_0 = A_0 \times B_0
\end{equation}
\begin{equation}\label{eq:karatsuba_1level_2}
T_2 = (A_0+A_1)\times(B_0+B_1)
\end{equation}
\begin{equation}\label{eq:karatsuba_1level_3}
\begin{gathered}
A\times B = \{A_1, A_0\} \times \{B_1, B_0\} \\= T_1 \times2 ^{N} + (T_2-T_1-T_0)\times2^{N/2}+T_0
\end{gathered}
\end{equation}

The Karatsuba algorithm can be implemented on hardware to exploit its reduced number of multiplications to lower area complexity, as previously shown in \cite{11_rafferty2017evaluation,  12_san2012increasing}, and many others. The Karatsuba algorithm presents some overhead to produce the subproducts, and in practice, it is usually beneficial for large multiplications, whereas Schoolbook multiplication remains more efficient for smaller ones.

While multiply-accumulate units have been extensively studied in the literature, the prior work lacks ASIC-focused multi-cycle multiplications.

Since MACs have similar flows to multi-cycle multipliers consisting of the accumulation of several multiplications, we can consider previous MAC-focused works as well since they can be adapted to become multi-cycle multipliers with some adjustments. \cite{MAC_07_HS_AE} proposes a high-speed, area-efficient MAC unit optimized for digital signal processing, leveraging parallelized partial product reduction. Their approach merges the accumulator with the compression tree, improving both area and speed. A similar approach can be applied to multi-cycle multipliers, where the subproducts can be fed back into the compression tree. More recent work in MAC architectures includes \cite{MAC_10_HS_AE_2C}, which introduces a two-cycle, energy-efficient design with double throughput, and \cite{MAC_20_HP_MA}, which integrates additions directly into partial product reduction to minimize latency. These works highlight the tradeoffs between throughput, area, and energy in multi-cycle arithmetic units.

Many architectures from MAC units can be adapted for multi-cycle multi-cycle multiplication. However, MAC-focused designs introduce some overhead from wider adders, accumulation registers, and other hardware for any additional features. In contrast, our work focuses on multi-cycle multiplication, eliminating such overheads and adapting the architectures to optimize area and energy further.

Several studies have been conducted to evaluate multi-cycle multipliers in the past \cite{11_rafferty2017evaluation, 12_san2012increasing, 26_von2005efficient, 27_gao2009efficient, 28_de2009large, 1_langhammer2021folded, Kumm, 19_li1996design}. These studies have mainly targeted FPGA applications, where the resources are more limited and require the design to efficiently accommodate the FPGA primitives (DSPs). However, in the case of ASICs, such design restrictions do not apply; thus, there is room for more efficient designs that leverage the customizability of ASICs more efficiently. For this reason, MCIMs are not directly comparable with these works.

\section{Proposed Architectures}\label{proposed architectures}
Multiplication can be expressed as a sum of several smaller multiplications using either the Schoolbook or Karatsuba approach. If the desired throughput is $<$ 1, resource sharing can be utilized to reduce the area complexity. Using complete multipliers for the smaller multiplications is inefficient in terms of timing due to the final addition. A better approach is to use Partial Product Multipliers (PPMs) for these smaller multiplications. PPMs are multipliers that omit the final addition (producing two results). The PPM can be designed based on existing multiplier architectures, such as Wallace and Dadda multipliers. However, Synopsys offers PPM generators \cite{synopsys} (DW02\_multp) that produced better results for our designs.  
Using a multi-cycle approach, the number of sub-products to be summed will be twice the Cycle-Time (CT), which can be efficiently accommodated using compressors. Similar to the PPMs, Synopsys also offers compressor generators \cite{synopsys} (DW02\_tree) which produced better results for most cases. However, for the feedforward architecture (which will be discussed in Subsection \ref{subsection:feed-forward}), a custom compressor that aims to minimize the area resources at the cost of wider final results produced improvements for some cases.  This approach can reduce the area complexity since the final addition can be spread into multiple cycles through the use of resource sharing or pipelining.

And finally, summing the results of the compressor produces the multiplication result.

\subsection{Feedback Architecture}\label{subsection:feedback}
The feedback (FB) architecture is based on the Schoolbook approach (distributive property) where one or both operands are divided to smaller parts. For CTs $\geq$ 4, it is possible to separate both multiplicands into smaller parts. However, separating only one variable maintains a lower critical path with no significant effect on the area complexity. 

\begin{figure}[!ht]
\begin{center}
\includegraphics[scale=0.65]{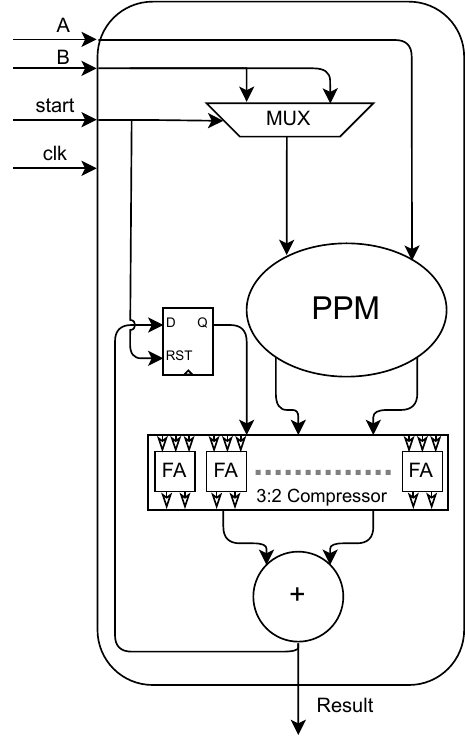}
\end{center}
\caption{ FeedBack (FB) architecure}
\label{fig:design_32}
\end{figure}

Dividing the multiplication into several smaller counterparts is straightforward; one operand can be divided into several smaller parts and then multiplexed to the PPM's input. The next step for the multiplication is to shift and sum the results of these multiplications. As we utilize a single PPM, we produce 2 results in every clock cycle. And since this is a multi-cycle multiplier, we also have a result from the previous clock cycle. Therefore, we have 3 results to sum from the previous cycle and 2 from the current cycle. Summing these directly with 2 adders is inefficient due to the long critical path. Thus, we can use a 3:2 compressor and a single adder. This approach maximizes the resource sharing that can be applied, which is due to the feedback loop. The feedback architecture is presented in Fig. \ref{fig:design_32}. The feedback loop connects the output of the adder back to the compressor. The critical path here is determined as the path of the adder and compressor. However, this also means that the design is expected to achieve lower maximum frequencies as the bit width increases since the adder cannot be pipelined due to the loop. This architecture is similar to the MAC units proposed in \cite{MAC_07_HS_AE}, \cite{MAC_DA}, and \cite{MAC_10_HS_AE_2C}, where compressors and feedback loops are utilized to reduce the critical path and area complexity.

\subsection{Feedforward Architecture}\label{subsection:feed-forward}
\begin{figure}[!ht]
\begin{center}
\includegraphics[scale=0.65]{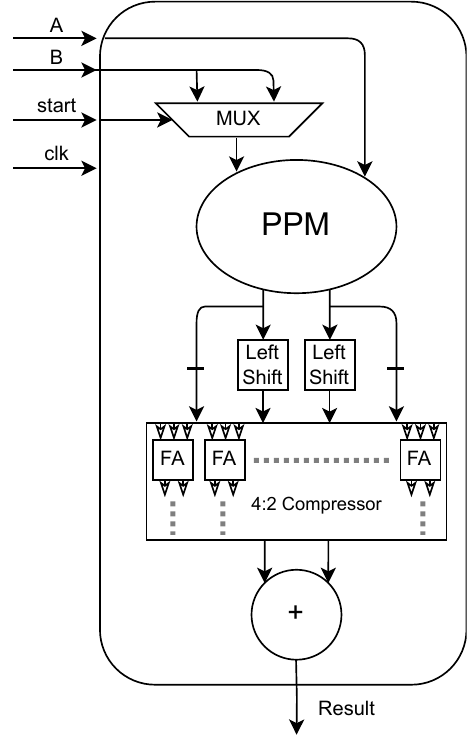}
\end{center}

\caption{Feedforward (FF) architecture}
\label{figure:desgin_42_2c_t}
\end{figure}

Multipliers are frequently used with high-frequency applications. Having a feedback loop in the design limits the design's ability to be pipelined. For this reason, we propose a design that contains no feedback loops, the feedforward (FF) architecture, allowing it to be pipelined to meet strict timing targets. This architecture is represented by Fig. \ref{figure:desgin_42_2c_t}. It first computes all the partial product multiplications using the PPM, storing the results in registers, and then sums the results using 4:2 compressor and final adder (while shifting accordingly). Such an architecture, however, is only area-efficient for a CT of 2 because the area complexity increases as CT is further increased. The compressor would have to handle more reductions, and more registers are required to hold the intermediate registers. 

\subsection{Karatsuba Architecture}\label{subsection:karatsuba}
\begin{figure}[!ht]
\begin{center}
\includegraphics[scale=0.7]{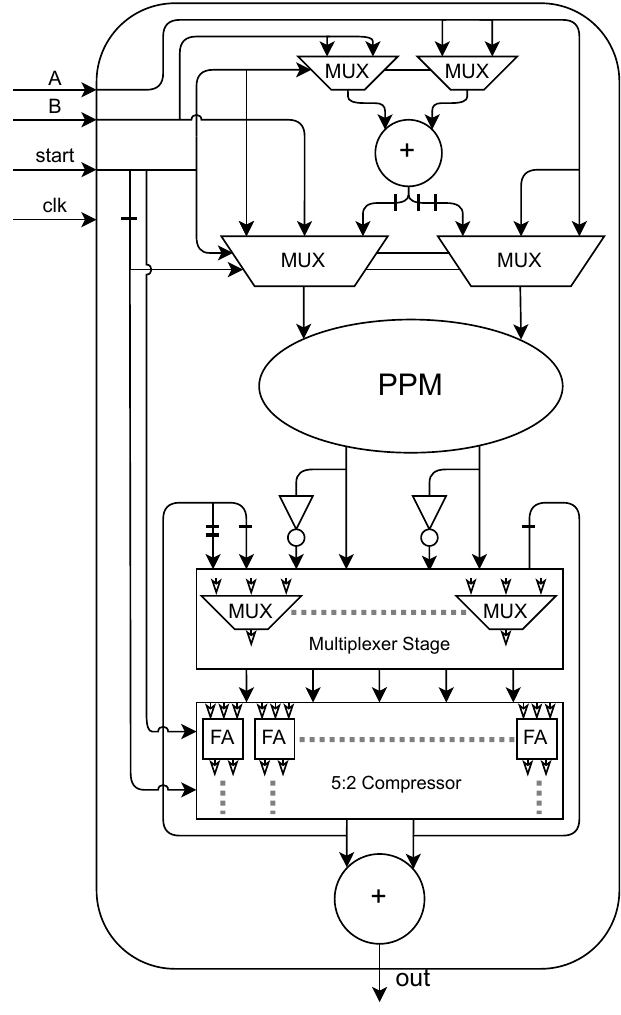}
\end{center}
\caption{ Karatsuba architecture}
\label{fig:design_karat_top}
\end{figure}

The Karatsuba approach is commonly used to create area-efficient multipliers. It allows can allow for smaller sub-multiplications at the expense of some overhead. This overhead, however, is outweighed by the area savings for larger multiplications. Fig. \ref{fig:design_karat_top} represents a Karatsuba design with a CT of 3. This architecture uses a small feedback loop around the compressor only, allowing for a smaller compressor and reducing the required resources while also allowing the final adder to be pipelined. The addition and subtraction operations from the Karatsuba algorithm are handled through the compressor by applying inverted inputs and constants, which eliminates the need for an additional adder to handle the two's complement.  

This architecture can be further optimized for large multiplications by applying the Karatsuba approach recursively to the PPM. Following this idea, a Karatsuba PPM consists of three smaller PPMs and a 10:2 compressor to handle the PPMs' outputs. Thus, any degree of recursion can be applied using this approach. Large multiplications can benefit from greater degrees of recursion, similar to how a combinational Karatsuba multiplier is expected to perform.

The feedforward and Karatsuba architectures compute the final addition separately. Omitting the final addition from these designs presents a new multi-cycle PPM. Multi-cycle PPM can be applied to the Karatsuba and feedforward architectures in a recursive fashion to achieve CTs that are factors of 2 and 3.

\subsection{Implementation Details}

To provide a performance benchmark, we compare our results with a standard multiplier (inferred by the "*" operator), which will be referred to as Star. The traditional approach for generating results is to set the design parameters, such as the CT and latency, and synthesize a design using the best achievable frequency. However, this approach is not the best representation of the results since the target frequency is usually a design decision or constraint. The area complexity, power consumption, and operating frequency are all interdependent. Thus, the timing target selection significantly affects the results. For this reason, we select the timing target first and then update a design's pipeline depth to meet the target timing. The same timing targets are used when comparing different designs. We also utilize the retiming feature (which is a technique used to optimize digital circuits by moving flip-flops) to achieve optimized pipelining. Registers are added at the end of the design to increase the depth of the pipeline. 

Our designs were verified by utilizing the Star multiplier as a reference and providing random test vectors to both Star and the respective MCIM multiplier, where the outputs of both should always match.

\section{Results} \label{results}

The proposed designs were tested thoroughly using different bit widths, latencies, and timing targets, utilizing the Synopsys Design Compiler and the TSMC 40nm technology.

This section will present tables that represent strict and relaxed timing conditions for both small and large multiplications. This work targets low CTs. Thus, the tables presented in subsections \ref{sec:syn_res_rel} and \ref{sec:syn_res_strict} focus on designs with CTs = 2 and 3. Additionally, subsection \ref{sec:syn_cts} presents the results for different CTs to show how the feedback design scales with lower throughputs.

\subsection{Synthesis Under Relaxed Timing Conditions}\label{sec:syn_res_rel}
Using a relaxed timing target presents the area complexity of each design without being affected by the critical path. The relaxed timing conditions case is represented by a timing target of 10 ns. Such a target allowed all the designs to meet timing without requiring additional pipeline stages while being able to use small library cells. Additional pipeline stages are not required since the timing target is very relaxed. Thus, the latencies presented in table \ref{table:syn_designs_relaxed} represent these designs' minimum latency (L). 

\begin{table}[!ht]
\caption{Synthesis Results for Under Relaxed Timing Conditions (Target = 10 $\mathrm{ns}$)}
\begin{center}
\begin{tabular}{| c | c | c | c | c | c |c |}  
\hline
 \multirow{2}{*}{Design} &\multirow{2}{*}{CT} &\multirow{2}{*}{Comp.} &\multirow{2}{*}{L} & \multirow{2}{*}{Area}& Power     & Ener.\\
        &     &       &    &      &  (mW)& (pJ)\\
 [0.5ex] 
 \hline
 \hline
 \multicolumn{7}{|c|}{ 16$\times$16 Multipliers} \\
 \hline
 Star & 1  & - & 1 & 1348  & 0.170 & 1.70\\ 
 \hline
 \multirow{2}{*}{FB}  & 2  & FAs & 2 & 942 & 0.105 & 2.10 \\ 
 \cline{2-7}
  &3 & FAs & 3 &  748 & 0.78 & 2.34\\ 
   \hline

 \multirow{2}{*}{FF} & 2  & DW02 & 2 & 1096 &   0.113&2.26\\ 
 \cline{2-7}
  &2  & Custom & 2 & 1105 &  0.113 &2.26 \\ 
 \hline
 \hline

 \multicolumn{7}{|c|}{ 128$\times$128 Multipliers} \\

 \hline
 Star &1   & - & 1 & 66319   &  11.38 & 113.8\\ 
 \hline
  \multirow{2}{*}{FF} &2   & DW02 & 2 &  37042 & 4.62 & 92.40\\ 
 \cline{2-7}
   &2  & Custom & 2 & 37790  & 6.15 &123.00\\ 
 \hline

 \multirow{2}{*}{FB} &2  & FAs & 2 & 42913  &  4.03 &80.60 \\ 
 \cline{2-7}

  &3 & FAs & 2 & 30217   & 3.36 & 100.80\\ 
 \hline

 \multirow{1}{*}{Karatsuba} &3    & DW02 & 3 & 31743 &  4.25 &127.50 \\ 
 \hline
\end{tabular}
\end{center}
\label{table:syn_designs_relaxed}
\end{table}
Karatsuba designs favor larger multiplications and do not provide area savings for small multiplications; therefore, the results are omitted for 16x16 multiplications as they do not offer significant area savings. As intended, the FeedBack (FB) designs best suit applications that do not require very strict timing targets. The FB designs can offer around 30\% area savings for a throughput of 1/2 and around 45\% for a throughput of 1/3 for 16$\times$16 multiplications. MCIM designs have a lower average peak power, which is the expected power consumption when inputs arrive back to back. Average peak power is an important consideration since it affects the required power delivery network and thermal performance. 

Area reductions are much more significant for large multipliers due to their large area. The Karatsuba MCIMs were specifically designed for large multiplications. Thus, they have an advantage when it comes to large multiplications. 

For 128$\times$128 multiplications, the feedback design with CT=3 consumed the least area of these designs, offering area savings of up to 54\%. The Karatsuba design offers similar savings and can outperform the feedback architecture for larger bit widths.

\subsection{Synthesis Under Strict Timing Conditions}\label{sec:syn_res_strict}
An important aspect of ASICs is their ability to operate at higher frequencies than the FPGA implementation counterpart. All MCIM designs were tested using strict timing conditions to demonstrate their ability to offer area savings for high-speed applications. The designs are tested using a timing target that cannot be met without pipelining. For 16$\times$16 multiplications, this is set as 0.31 ns, which is equal to the clock-to-q + setup + hold delay for 1 FA placed between registers. For larger 128$\times$ multiplications, 0.31 ns would require too many pipeline stages. Thus, we selected a more reasonable timing target of 0.8 ns based on the observed timing results. Table \ref{table:syn_designs_strict} presents the synthesis results under strict timing conditions. 

\begin{table}[!ht]
\caption{Synthesis Results Under Strict Timing Conditions }
\begin{center}
\scalebox{0.98}{
\begin{tabular}{|c | c| c | c | c | c | c | c |  c |}  
\hline

\multirow{2}{*}{Design}	 &\multirow{2}{*}{CT} & \multirow{2}{*}{Comp.} & \multirow{2}{*}{L} & \multirow{2}{*}{Area} & Slack   &Pow. & Ener.\\ [0.5ex] 
      &       &        &         &      &
     (ns) & (mW) &(pJ)\\[0.5ex] 
 \hline
 \hline
  \multicolumn{8}{|c|}{ 16$\times$16 Multipliers, Timing Target = 0.31 ns} \\
 \hline
 Star & -  & - &  7 & 5178  & 0 & 13.34 & 4.14 \\ 
 \hline

  \multirow{2}{*}{FF} & 2   & DW02 & 9 & 3984 & 0 & 6.76 &4.19 \\ 
 \cline{2-8}

 \cline{2-8}
   & 2   & Custom & 9 & 4065 & 0 & 7.18 & 4.45\\ 
 \hline

 \multirow{2}{*}{FB} & 2   & FAs & 4 & 3712 &  -0.15 &  6.58&6.05\\ 

 \cline{2-8}
  &3   & FAs & 9 & 3625 & -0.15  & 6.14 & 8.47 \\ 

 \hline
  \hline

  \multicolumn{8}{|c|}{ 128$\times$128 Multipliers, Timing Target = 0.8 ns} \\
 \hline
 Star & 1 & -& 4 & { 121634}  & 0 & 98.14& 78.51 \\ 
 \hline
 
  \multirow{2}{*}{FF} &2   & DW02 & 4 & 67884 & 0  & 33.01 & 52.82  \\ 
 \cline{2-8}
 &2    & Custom  & 4 & 64778  & 0 & 31.83 & 50.93  \\ 
 \hline
  \multirow{2}{*}{FB}& 2   & FAs  & 4 & 92240 & 0 & 47.47 & 75.95\\ 
 \cline{2-8}
 & 3  & FAs  & 6  & 48068 & 0  & 30.45 & 73.08\\ 
 \cline{1-8}

  \multirow{1}{*}{Karatsuba} &3    & DW02  & 7 & 44888 & 0  & 26.95 & 64.68\\ 
 \hline
\end{tabular}}
\end{center}
\label{table:syn_designs_strict}
\end{table}

For smaller 16$\times$16 multiplications, the feedforward design is the only design that can meet such a timing target. This is due to its ability to be efficiently pipelined, offering up to 23\% area savings. The feedback design was not able to meet the timing target due to the feedback loop, which limits how effectively the design can be pipelined. Similar to the case of relaxed timing conditions, MCIM designs maintain a lower average peak power.

For larger 128$\times$128 multiplications, Karatsuba designs are the most efficient multipliers as they offer the most area savings. The Karatsuba designs offer area savings of 63\% as well as energy savings of 18\%. The feedforward design that uses the DW02 PPM and the Custom compressor offers area savings of 47\% and energy savings of 33\%. When MCIM designs offer such savings, even utilizing two 3-cycle designs becomes viable; such a combination can achieve a throughput of 2/3 while offering an area saving of up to 26\%. Additionally, a combination of one 2-cycle multiplier and one 3-cycle is also viable, offering a combined throughput of 5/6 and an area saving of up to 9\%. Depending on the application's requirements, which include energy consumption, area limitation, and throughput requirement, any of these designs or a combination of these designs can be the optimal choice.

\subsection{Results for Other Cycle Times} \label{sec:syn_cts}

The architectures presented in this work are suitable for any CT. However, the feedback design is the most appropriate design of our architectures for CT $\geq$ 3. The feedforward design is only area-efficient for a CT of 2 since increasing the CT impacts the area negatively. The Karatsuba algorithm gives optimal results for a CT of 3. However, it can be implemented for any CT that is a multiple of 3. This, nevertheless, is only beneficial for very large multiplications. The feedback design scales well as the CT increases, offering an increase in area savings and a reduction in average peak power. Table \ref{table:differentCTs} includes the synthesis results of 32 $\times$32 bit multipliers using different CTs (2-8). Feedback designs scale well for greater CTs, offering greater area reductions and reducing the average peak power.
\begin{table}[!ht]
\caption{Synthesis Results for 32 $\times$32 Multipliers\\ with CTs up to 8 (target = 10 $\mathrm{ns}$)}
\begin{center}\scalebox{1}{
\addtolength{\leftskip} {-0.5cm} 
\addtolength{\rightskip}{-0.5cm}
\begin{tabular}{ | c | c | c | c |c | c | }  
\hline
 \multirow{2}{*}{Design}	 & \multirow{2}{*}{CT} &
 \multirow{2}{*}{Area} &
 Area 
 & Power
 & Energy   \\ [0.5ex] 
 
   & & &Savings & (mW) &   (pJ)   \\ [0.5ex] 
 \hline
 \hline
 Star & 1  & 4349 & - & 0.74 & 7.4\\
  \hline
  \multirow{7}{*}{FB} &  \multirow{1}{*}{2}  & 2624 & 40\% & 0.46 & 9.2 \\
   \cline{2-6}

& \multirow{1}{*}{3}  & 2188 & 50\% & 0.29 & 8.7\\
   \cline{2-6}

  & \multirow{1}{*}{4}  & 1876 & 57\% &0.23  & 9.2\\
   \cline{2-6}

  & \multirow{1}{*}{5}  & 1739 & 60\% & 0.17& 8.5\\
   \cline{2-6}

  & \multirow{1}{*}{6}   & 1567 & 64\% & 0.15 & 9.0\\
   \cline{2-6}

  & \multirow{1}{*}{7}  & 1385 & 68\% & 0.14 & 9.8\\
  \cline{2-6}

   & \multirow{1}{*}{8}   & 1206 & 72\% & 0.12 & 9.6\\

  \hline
  
\end{tabular}
}\end{center}
\label{table:differentCTs}
\end{table}

\subsection{Discussions}
The proposed MCIM designs can offer significant area savings for various applications. The feedforward design is best suited for strict timing targets, the feedback design is best suited for more relaxed timing targets, and the Karatsuba design is best suited for multiplications of sizes 128 and greater.

A combination of two MCIM designs is also possible. A combination of 2- and 3-cycle designs can achieve a throughput of 5/6, and a combination of two 3-cycle designs achieves a throughput of 2/3. These combinations can still offer area savings for large multipliers since the combined area of these designs is less area than a Star multiplier. 

\subsection{Use Cases}\label{use cases}
MCIM designs can be used in two main cases. These cases rely on the required throughput, latency constraints, and target timing.

\subsubsection*{Case 1: Required throughput is $<1$}
When $i$ multiplications are required within $j$ cycles, and $i/j$ is not an integer. A single MCIM design can be used to offer area savings. For example, if an application requires a throughput of 3/2, a single Star multiplier (throughput=1) can be used with any of our 2-cycle designs (throughput = 1/2), assuming there are no latency constraints. Such cases can be found in applications with inter-operation dependencies and other design specifications. MCIM designs can require longer latencies than Star to be area-efficient, usually 1 or 2 additional cycles. So latency constraints should be evaluated when considering MCIM designs.

\subsubsection*{Case 2: Latency constraint and strict timing target}
\begin{figure}[!ht]
\begin{center}
\includegraphics[scale=1]{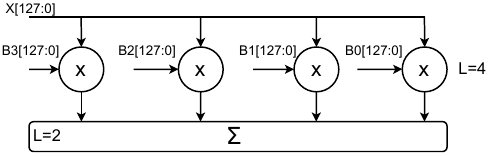}
\end{center}
\caption{ FIR filter with a latency constraint of 6 cycles of 0.8 ns}
\label{fig:fir_filter}
\end{figure}

The use of MCIM designs can be applicable in scenarios where a latency constraint is required. For instance, when 4 multiplications need to be executed within 4 cycles, such as in the example given in Fig. \ref{fig:fir_filter}, the only viable options to meet the timing target of 0.8 ns are the Star multiplier with a latency of 4 (1C4L) and the feedforward multiplier with a latency of 4 (2C4L). While using a single 1C4L multiplier or 2 1C4L designs may be more area-efficient, they both violate the latency constraint. Therefore, the only way to meet the latency constraint is to use 4 individual multipliers in parallel. The 2C4L multiplier offers an area reduction of 47\%, making it the most area-efficient solution for this case when 4 multipliers will be used in parallel (if $throughput < 0.5$).

\section{Conclusion} \label{conclusion}

Fast combinational multipliers can consume significant silicon area power, especially for large-bit widths. The area can be minimized by implementing resource sharing (such as folding), which may be suitable for applications that do not require an output every cycle.

Although FPGA-focused multi-cycle multipliers and multiply-accumulate units have been widely researched, previous studies do not address ASIC-specific multi-cycle multiplications. This work demonstrates the benefits of multi-cycle multiplication for the area and energy efficiency as opposed to using single-cycle multipliers, which would, otherwise, be underutilized. 

We propose the feedback, feedforward, and Karatsuba Multi-Cycle Integer Multipliers (MCIM), which are optimized for different application requirements. The feedback architecture is often the optimal choice for applications running that do not require very high frequencies and can be used with any CT (Cycle-Time) $\geq$2. The feedforward architecture is optimal for high-frequency applications with a CT of 2. The Karatsuba architecture, optimized for large multiplications, is often the optimal choice for bit width $\geq 128 \times 128$ and is suitable for both low- and high-frequency applications with a CT of 3.

The proposed MCIM designs with a CT of 2 can offer up to 44\% area reductions and 33\% energy savings with respect to a standard multiplication using the * operator. And can offer greater area savings for larger CTs. Overall, these designs can offer significant area and power savings which are specifically important for edge and IoT devices.

Future research could explore alternative architectures inspired by state-of-the-art MAC implementations. Additionally, extending the design to support variable CTs could enable flexible operation, such as performing  N×N multiplications in one cycle and 2N×2N in the following four cycles, improving adaptability to different workloads. Furthermore, the architectures can be extended to support signed multiplication to support a wider range of applications.

\bibliographystyle{IEEEtran}
\bibliography{output.bib}
\end{document}